\documentclass[prl,aps,floats,superscriptaddress,amsfonts,twocolumn,showpacs]{revtex4}

\usepackage{graphicx}

\begin{document}

\title{Proposal for a loophole-free
Bell test using homodyne detection}

\author{R.~Garc\'{\i}a-Patr\'{o}n S\'{a}nchez}
\affiliation{QUIC, Ecole Polytechnique, CP 165,
Universit\'e Libre de Bruxelles, 1050 Bruxelles, Belgium}

\author{J. Fiur\'a\v{s}ek}
\affiliation{QUIC, Ecole Polytechnique, CP 165,
Universit\'e Libre de Bruxelles, 1050 Bruxelles, Belgium}
\affiliation{Department of Optics, Palack\'y
University, 17. listopadu 50, 77200 Olomouc, Czech Republic}

\author{N.J. Cerf}
\affiliation{QUIC, Ecole Polytechnique, CP 165,
Universit\'e Libre de Bruxelles, 1050 Bruxelles, Belgium}

\author{J. Wenger}
\affiliation{Laboratoire Charles Fabry de l'Institut
d'Optique, CNRS UMR 8501, F-91403 Orsay, France}

\author{R. Tualle-Brouri}
\affiliation{Laboratoire Charles Fabry de l'Institut
d'Optique, CNRS UMR 8501, F-91403 Orsay, France}

\author{Ph. Grangier}
\affiliation{Laboratoire Charles Fabry de l'Institut
d'Optique, CNRS UMR 8501, F-91403 Orsay, France}

\begin{abstract}
We propose a feasible optical setup allowing for a loophole-free Bell
test with efficient homodyne detection. A non-gaussian entangled state
is generated from a two-mode squeezed vacuum by subtracting a single
photon from each mode, using beamsplitters and standard low-efficiency
single-photon detectors. 
A Bell violation exceeding 1\% is achievable with 6~dB squeezed
light and an homodyne efficiency around 95\%. A detailed feasibility analysis,
based upon the recent generation of single-mode non-gaussian states,
confirms that this method opens a promising avenue towards a complete
experimental Bell test.
\end{abstract}


\maketitle

In their seminal 1935 paper, Einstein, Podolsky, and Rosen (EPR)
advocated that if ``local realism'' is taken for granted, then quantum theory is
an incomplete description of the physical world \cite{Einstein35}.
The EPR argument gained a renewed attention in 1964
when John Bell derived his famous inequalities, which must be satisfied
within the framework of any local realistic theory \cite{Bell64}.
The violation of Bell inequalities (BI), predicted by quantum mechanics,
has since then been observed in many experiments
\cite{Freedman72,Aspect81,Weihs98,Tittel98,Rowe01},
disproving the concept of local realism.
So far, however, all these tests suffered from ``loopholes'' allowing
a local realistic explanation of the experimental observations
by exploiting either the low detector efficiency \cite{Pearle70}
or the time-like interval between the two detection events
\cite{Santos92}. In this Letter, we propose a realistic experimental
scheme based on the conditional generation of non-gaussian entangled 
light states and balanced homodyning, which can circumvent these difficulties 
and allows for a ``loophole-free'' experimental Bell test.
\par

A test of Bell inequality violation typically involves two distant
parties, Alice and Bob, who simultaneously carry out measurements on two
systems prepared in an entangled quantum state.
The measurement events (including the measurement choice)
at Alice's and Bob's sites must be spacelike separated
in order to avoid the locality loophole, that is,
to rule out any possible communication between the two parties.
Optical systems are therefore particularly suitable candidates for Bell
tests, because entangled photon pairs can now be generated
and distributed over long distances \cite{Weihs98,Tittel98}.
However, the currently available single-photon detectors suffer
from too low efficiencies $\eta$, opening the so-called
detector-efficiency loophole, that is, the experimental data
can be explained by local realistic theories wherein the detectors
only click with probability $\eta$. This loophole is
present in all optical BI tests today.
\par

In contrast,
very high detection efficiencies can be reached in optical systems
using balanced homodyne detectors \cite{Polzik92,zhang03,Grosshans03},
opening a very promising alternative to Bell tests based on
single-photon detectors. Several theoretical works have shown that 
a violation of BI
may indeed be observed with balanced homodyning provided
that some very specific entangled light states can be prepared
\cite{Gilchrist98,Munro99,Wenger03}.
Although a violation up to the maximum theoretical
limit can be achieved with homodyne detection \cite{Wenger03},
all the states required in \cite{Gilchrist98,Munro99,Wenger03}
unfortunately appear to be experimentally
infeasible. On the other hand, the experimentally accessible
\cite{Ou92,Schori02,Bowen04} two-mode squeezed vacuum states are unsuitable 
for a Bell test because they are characterized by a positive-definite 
gaussian Wigner function, which provides an explicit hidden variable model 
for homodyne measurements.
So far, no feasible experimental scheme has been found
that could be used to prepare non-gaussian states
that exhibit a violation of BI with balanced homodyning.
\par

In this Letter, we show that a strikingly simple optical setup (see Fig.
1) can be used to conditionally generate non-gaussian states that are
suitable for this purpose. Our scheme requires a pulsed source of two-mode 
squeezed vacuum state, which can be expressed in the Fock basis as
\begin{equation}
|\psi_{\mathrm{in}}\rangle_{AB}
=\sqrt{1-\lambda^2}\sum_{n=0}^\infty \lambda^n
|n,n\rangle_{AB} ,
\label{psiin}
\end{equation}
where $\lambda=\tanh(r)$ and $r$ is the squeezing constant.
The basic idea is to ``degaussify''
this state by subtracting a photon from each mode
\cite{Opatrny00,Cochrane02,Olivares03}.
More precisely, we produce a non-gaussian entangled state
with the use of two unbalanced beam splitters BS$_A$ and BS$_B$
with intensity transmittance $T$,
followed by two photon-counting detectors
PD$_A$ and PD$_B$ such as avalanche photodiodes (APD). The
successful state preparation is heralded by a click of both
PD$_A$ and PD$_B$. We shall see that a detector efficiency
as low as $\eta=10$\% suffices for the preparation of states
exhibiting a violation of BI. 
Very recently, the generation of pulsed single-mode non-gaussian states
by photon subtraction from squeezed states
has been demonstrated experimentally by some of us \cite{WBG}.
We therefore can realistically envision the experimental realization
of a Bell test with non-gaussian two-mode states prepared along the same
lines.
\par

\begin{figure}[!t!]
\centerline{ \includegraphics[width=6cm]{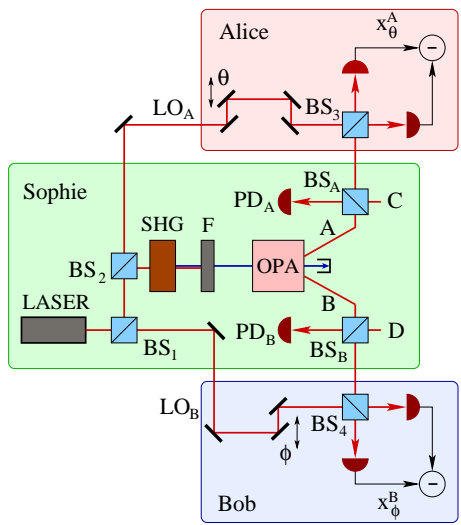} }
\caption{Proposed experimental setup. The source (controlled by
Sophie) is based on a master laser beam, which is
 converted into second harmonic in a nonlinear
crystal (SHG). After spectral filtering (F), the second harmonic beam
pumps an optical parametric amplifier (OPA)
which generates two-mode squeezed vacuum in modes A and B. Single
photons are conditionally subtracted from modes A and B with the use of the beam
splitters BS$_A$ and BS$_B$ and single-photon detectors PD$_A$ and
PD$_B$. Alice (Bob) measures a quadrature of mode A (B) using a balanced
homodyne detector that consists of a balanced beam splitter BS$_3$ (BS$_4$) and a pair of highly-efficient photodiodes. The local
oscillators LO$_A$ and LO$_B$ are extracted from the laser beam by means of
two additional beam splitters BS$_1$ and BS$_2$.}
\end{figure}

The photon subtraction can easily be understood
in the limit of high transmittance of BS$_A$ and BS$_B$ ($T\rightarrow
1$), since then the most probable event leading to a click at each PD
is when a single photon is reflected by each BS. In this limit, the
conditionally generated state is very close to a pure
state~\cite{Cochrane02}
\begin{equation}
|\psi_{\mathrm{out}}\rangle_{AB} \propto \hat a_A \hat a_B
|\psi_{\mathrm{in}}\rangle_{AB}
\propto \sum_{n=0}^\infty (n+1)(T\lambda)^n|n,n\rangle_{AB}  ,
\end{equation}
where $\hat a_{A,B}$ are annihilation operators. This state is non-gaussian,
and therefore suitable for a Bell test based on homodyning.
\par

In the experiment proposed here, Alice and Bob measure the quadratures
$x_{\theta}^A= \cos\theta \, x^A+\sin\theta \, p^A$ and
$x_{\phi}^B=\cos\phi \, x^B+\sin\phi \, p^B$, which have continuous
spectra and satisfy $[x^j,p^k]=i\delta_{jk}$.
In order to use the standard Bell-CHSH inequality \cite{CHSH},
we discretize these quadratures by postulating
that the outcome is $+1$ when $x\geq 0$, and $-1$ otherwise.
Alice and Bob must then choose randomly and independently between two different 
measurements, corresponding to the choices of two angles $\theta_1,\theta_2$
and $\phi_1,\phi_2$. Their measurements outcomes are thus described by four 
variables $a_1,a_2,b_1,b_2$, with values $+1$ or $-1$.
One defines the usual Bell parameter
\begin{equation}
S=\langle a_1 b_1\rangle+\langle a_1b_2\rangle+\langle
a_2b_1\rangle-\langle a_2
b_2\rangle,
\label{S}
\end{equation}
where $\langle a_j b_k\rangle$ denotes the average over the subset of
the experimental data when Alice measured $a_j$ and, simultaneously,
Bob measured $b_k$. As we shall see, exploiting the fact
that PD$_A$ and PD$_B$ can be viewed here as
``event-ready'' detectors \cite{Bell}, one can
prove that all local-realistic models for
Alice and Bob measurements must satisfy the Bell-CHSH inequality
$|S|\le 2$ \cite{CHSH}. 
\par

In the formalism of ``event-ready" detectors
introduced by John Bell \cite{Bell},
one should know, by some initiating event, when a measurable system
has been produced. The main idea is to pre-select
-- rather than post-select -- the relevant events. For that purpose,
one considers 3 partners, Alice and Bob who perform the measurements,
and Sophie who controls the source, see Fig.~1.
The entire data analysis must then be performed on a
pulsed basis, with Sophie sending time-tagged light pulses
(local oscillator and squeezed light) to Alice and Bob.
In each experimental run, Sophie records whether
her photodetectors clicked, while Alice and Bob carry out measurements
of one of two randomly chosen quadratures.
After registering a large number of events,
the three partners discard all events not corresponding to an ``event-ready"
double-click registered by Sophie.
The correlation coefficients $\langle
a_j b_k \rangle$ are then evaluated from all remaining events,
and plugged into the $S$ parameter (\ref{S}).
\par

In a local realistic approach, the light pulses in each time slot
carry some random unknown parameters $\mu$, which will ultimately
determine the sign of $x_{\theta}^A$ and $x_{\phi}^B$.
The crucial feature
is that the experiment should be designed in such a way that the
measurements at Alice's and Bob's sites, as well as the detection of
clicks from the conditioning detectors by Sophie, are all space-like
separated \cite{Aspect76}. Therefore, the probability distribution $p(\mu)$
must be independent of the measurement phases $\theta_{1,2}$ and
$\phi_{1,2}$.
The measured sign $s_A$ on Alice's side (resp. $s_B$ on Bob's side)
only depends on $\mu$ and $\theta$ (resp. $\phi$ on Bob's side).
One can thus
write $ \langle a_j b_k \rangle = \int d\mu \; p(\mu) \, s_A(\theta_j, \mu)
\, s_B(\phi_k, \mu)$, from which the derivation of the Bell-CHSH inequality 
is very standard \cite{CHSH}. 
The only requirement needed in the above reasoning is local realism,
the absence of any other assumption being the condition
for a ``loophole-free'' Bell test.
\par

Let us now evaluate the quantum mechanical value of the
$S$ parameter in the proposed experimental scheme.  We only briefly
outline the main steps of the calculation, the details will be presented
elsewhere \cite{Sanchez04}. We model realistic photodetectors (which have
a limited quantum efficiency $\eta<1$, and cannot discriminate between
one and multiple photon detection events) as ideal detectors preceded
with a ``virtual'' beam splitter of transmittance $\eta$.
Ideal detectors respond with two different outcomes, either a
no-click (projection onto the vacuum) or a click (projection on all
states with at least one photon). Similarly, a balanced homodyne detector with
efficiency  $\eta_{\mathrm{BHD}}$ is modeled as a perfect homodyne
detector preceded with a ``virtual'' beam splitter
of transmittance $\eta_{\mathrm{BHD}}$.
At the output of the OPA, the modes A and B are prepared in a two-mode
squeezed vacuum state, and the auxiliary modes C and D are in vacuum
state. The Wigner function of the state of modes ABCD is
a gaussian function centered at the origin,
\begin{equation}
W_{\mathrm{in},ABCD}(r)=\frac{1}{\pi^4
\,\sqrt{\det\gamma_{\mathrm{in}}}}
\exp\left[-r^T \gamma_{\mathrm{in}}^{-1}r\right].
\label{Wignerin}
\end{equation}
This state is fully characterized by the covariance matrix
$\gamma_{ij}= \langle r_i r_j+r_j r_i\rangle-2\langle r_i\rangle\langle
r_j\rangle$,
where $r=(x^A,p^A,\ldots,x^D,p^D)$ is a vector of quadrature components.
The input covariance matrix is
$\gamma_{\mathrm{in}}=\gamma_{\mathrm{in},AB}\oplus I_{CD}$,
where  $I$ is the identity matrix
and $\gamma_{\mathrm{in},AB}$ denotes the covariance matrix of the
two-mode squeezed vacuum (\ref{psiin}).
After combining modes A and C (B and D) on an unbalanced beam
splitter BS$_{A}$ (BS$_B$) with transmittance $T$, the modes C and D are
detected by the photon-counting detectors PD$_A$ and PD$_B$, 
while modes A and B are sent to the homodyne detectors. 
The covariance matrix
$\gamma_{\mathrm{out}}$ of the mixed gaussian state
of modes ABCD after passing through the beam splitters BS$_A$ and BS$_B$
and the four ``virtual'' beam splitters modeling the imperfect detectors
is related to $\gamma_{\mathrm{in}}$ via
a gaussian completely positive map \cite{Eisert02},
$\gamma_{\mathrm{out}}=S_{\eta}S_{\mathrm{BS}}\gamma_{\mathrm{in}}
S_{\mathrm{BS}}^T S_{\eta}^T+G$,
where $S_{\mathrm{BS}}$ describes the mode coupling in BS$_A$
and BS$_B$, and the matrices
$S_{\eta}=\sqrt{\eta_{\mathrm{BHD}}}I_{AB}\oplus \sqrt{\eta}I_{CD}$ and
$G=(1-\eta_{\mathrm{BHD}})I_{AB}\oplus (1-\eta)I_{CD}$ account for
the imperfect detectors.
\par

The Wigner function of the state of modes A and B prepared by conditioning
on observing clicks at PD$_A$ and PD$_B$ 
can then be expressed as a linear combination of four gaussian functions,
\begin{equation}
W(r)= \frac{1}{\pi^2 P \sqrt{\det\gamma_{\mathrm{out}}} } \sum_{j=1}^4
\frac{q_j}{\sqrt{\det \Gamma_{j,CD}}} \exp[-r^T \Gamma_j r],
\label{Wignercond}
\end{equation}
where $P=(\det\gamma_{\mathrm{out}})^{-1/2}
\sum_{j=1}^4 q_j[\det (\Gamma_{j}\Gamma_{j,CD})]^{-1/2}$
stands for the probability of  success,
$r=(x^A,p^A,x^B,p^B)$, and we have defined $q_1=1$, $q_2=q_3=-2$,
$q_4=4$. The various matrices appearing
in Eq. (\ref{Wignercond}) are obtained from
$\gamma_{\mathrm{out}}^{-1}$, which can be split into four
smaller submatrices as
$\gamma_{\mathrm{out}}^{-1}=\left[
\begin{array}{cc}
\Gamma_{AB} & \sigma \\
\sigma^T & \Gamma_{CD}
\end{array}
\right]$.
These four submatrices are then used to define
$\Gamma_{j}=\Gamma_{AB}-\sigma \Gamma_{j,CD}^{-1}\sigma^T$,
where $\Gamma_{1,CD}=\Gamma_{CD},$
$\Gamma_{2,CD}=\Gamma_{CD}+I_{C}\oplus 0_D,$
$\Gamma_{3,CD}=\Gamma_{CD}+0_{C}\oplus I_D,$ and
$\Gamma_{4,CD}=\Gamma_{CD}+I_{CD}.$
\par

After discretization of the quadratures, the correlation coefficient
$E(\theta_j,\phi_k)\equiv\langle a_j b_k\rangle$
can be expressed as
\begin{equation}
E(\theta_j,\phi_k)=\int_{-\infty}^\infty \mathrm{sign}(x_{\theta_j}^A
x_{\phi_k}^B)
P(x_{\theta_{j}}^A,x_{\phi_{k}}^B) d x_{\theta_{j}}^A d x_{\phi_{k}}^B,
\label{E}
\end{equation}
where $P(x_{\theta_j}^A,x_{\phi_k}^B)$ is the joint probability
distribution of the two commuting quadratures $x_{\theta_{j}}^A$ and
$x_{\phi_{k}}^B$, which can be determined as a marginal distribution
from the Wigner function given by Eq. (\ref{Wignercond}).
The correlation coefficient $E$ (and therefore $S$)
can then be expressed in a closed form by analytically
integrating the resulting gaussian functions.
\par

\begin{figure}[!t!]
\begin{center}
\includegraphics[width=8.5cm]{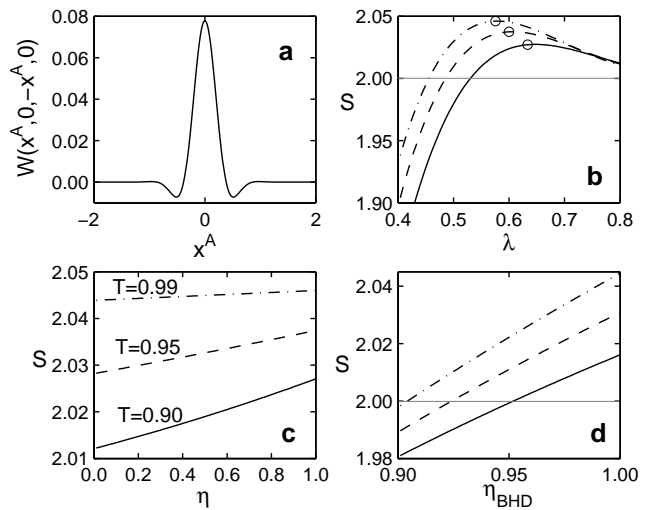}
\end{center}
\caption{Violation of Bell-CHSH inequality with the
conditionally-prepared non-gaussian state.
(a) One-dimensional cut of the Wigner function of the conditionally
generated non-gaussian two-mode state (with $\lambda=0.5$, $T=0.95$,
and $\eta=30$\%) along the line $x^B=-x^A$, $p^A=p^B=0$.
Notice the regions where $W$ is negative.
(b) Bell parameter $S$ as a function of the
squeezing $\lambda$ of the initial two-mode squeezed vacuum.
The curves are plotted for perfect detectors
($\eta=\eta_{\mathrm{BHD}}=100$\%)
with $T=0.9$ (solid line), $T=0.95$ (dashed line),
and $T=0.99$ (dot-dashed line).
The open circles mark the points where $T\lambda=0.57$.
(c) Bell parameter $S$ as a function of the efficiency
$\eta$ of the single-photon detectors, for $\lambda T=0.57$,
$\eta_{\mathrm{BHD}}=100$\%, and the same three $T$'s as in
(b). (d) Bell parameter $S$ as a function of the efficiency
$\eta_{\mathrm{BHD}}$ of the balanced homodyne detectors,
for $\lambda T=0.57$, $\eta=30$\%,
and the same three $T$'s as in (b).}
\end{figure}

The results plotted in Fig.~2 were obtained for the optimal choice of
angles $\theta_1=0$, $\theta_2=\pi/2$, $\phi_1=-\pi/4$, $\phi_{2}=\pi/4$.
Figure~2(a) shows that the Wigner function $W$
is negative in some regions of the phase space, which is
a necessary condition for the observation of a violation of BI 
with homodyne detectors. Fig.~2(b) illustrates that
the Bell inequality $|S|\le 2$ can indeed be violated with the proposed
set-up, and shows that there is an optimal squeezing $\lambda_{\mathrm{opt}}$
which maximizes $S$. A simplified calculation assuming perfect
detectors with single-photon resolution \cite{Sanchez04} predicts that
$\lambda_{\mathrm{opt}}T\approx 0.57$.
It follows from Fig.~2(c) that the Bell factor $S$ depends only very
weakly on the efficiency $\eta$ of the single-photon detectors, so the Bell
inequality can be violated even if $\eta<10$\%. In contrast, the efficiency 
of the homodyne detectors $\eta_{\mathrm{BHD}}$ 
must be above $\sim 90$\%, see Fig.~2(d).
\par

The maximum Bell factor achievable with our scheme
is about $S_{\mathrm{max}}\approx 2.046$ which represents a BI violation
of $2.3$\%. This is a small, but statistically
significant violation, which should be possible to observe experimentally.
The required degree of squeezing to get close to $S_{\mathrm{max}}$
corresponds to $\lambda\approx 0.57$, i.e., approximately 5.6~dB,
a value that has already been achieved experimentally
\cite{Polzik92,Lam99}. Another important parameter
is the transmittance $T$ of the beam splitters BS$_A$ and BS$_B$,
which must be as high as possible for maximizing $S$.
However, the probability of successful state preparation can be estimated
as $P\approx \eta^2 (1-T)^2$, so that it quickly drops when $T$ approaches 
unity. Thus there is a tradeoff between $S$ and $P$ which needs to be optimized
by taking into account the statistical uncertainties of the data.
\par

In order to be more specific, let us consider the single-mode version
of our scheme, which has already been implemented experimentally \cite{WBG}.
It is based on a commercial cavity-dumped titanium-sapphire laser,
delivering nearly Fourier-limited pulses at 850~nm,
with a duration of 150~fs and a repetition rate of 790~kHz.
Squeezed vacuum pulses, generated by parametric deamplification,
are sent through a beam splitter, and the reflected beam
is detected by a silicon APD. Conditional on observing a click,
the transmitted pulse is prepared in a single-mode non-gaussian state,
which is measured by homodyne detector with overall efficiency
$\eta_{\mathrm{BHD}}\approx 75$\%.
This experiment gives us useful estimates for a possible Bell test.
First, the delay between pulses ($1.2 \; \mu$s) allows
ample time for individual pulse analysis.
A fast random choice of the analyzed quadratures can be performed using
electro-optical modulators on the LO beams, triggered for instance
by digitizing the shot-noise of locally generated auxiliary beams.
Switching times around 100 ns, associated with propagation
distances of a few tens of meters, seem quite feasible.
The APDs can be triggered only when a pulse is expected,
reducing the effect of dark counts to a negligible value.
The intrinsic APD efficiency is about $50$\%, but
the filtering used to select a single mode currently
reduces the overall $\eta$ to less than $5$\%,
which should be improved for accumulating enough statistics.
\par

We can thus define a set of realistic
parameter values that should be reached in a loophole-free Bell test~:
with $\eta=30$\%, $T=95$\%, and $\lambda=0.6$,
BI are violated by about $1$\% if the homodyne efficiency 
$\eta_{\mathrm{BHD}}$ is larger than $95$\% (see Fig. 2(d)).
With a repetition rate of $1$~MHz and $P\approx 2.6 \times 10^{-4}$,
the number of data samples would be several hundreds per second,
so that the required statistics to see a violation in the percent range
could be obtained in a reasonable time (less than one hour).
In addition, the electronic noise of the homodyne detectors
should be 15-20 dB below shot noise, attainable with low-noise charge 
amplifiers. All these numbers have already been reached separately
in various experiments, but attaining them simultaneously certainly
represents a serious challenge.
Nevertheless, taking into account many possible experimental
improvements, the existence of an experimental window
for a loophole-free test of Bell inequalities can be considered
as highly plausible.
As a conclusion, it appears that, with quantum
continuous variables, a reasonable compromise can be found
between the experimental constraints
and the very stringent requirements imposed by a loophole-free test
of Bell inequalities.

\par

This work is supported by the European IST/FET program,
the Communaut\'e Fran\c{c}aise de Belgique under Grant No. ARC 00/05-251,
and the IUAP programme of the Belgian government under Grant No. V-18.
JF also acknowledges support from
the Grant No. LN00A015 of the Czech Ministry of Education.


\vspace{-0.5cm}
\begin{thebibliography}{99}
\vspace{-0.2cm}

\bibitem{Einstein35}
A. Einstein, B. Podolsky, and N. Rosen,
Phys. Rev. \textbf{47}, 777 (1935).

\bibitem{Bell64}
J.S. Bell, Physics (Long Island City, N.Y.) \textbf{1}, 195 (1964).

\bibitem{Freedman72}
S.J. Freedman and J.F. Clauser,
Phys. Rev. Lett. \textbf{28}, 938 (1972).


\bibitem{Aspect81}
A. Aspect, P. Grangier, and G. Roger,
Phys. Rev. Lett. \textbf{47}, 460 (1981);
{\it ibid.}, \textbf{49}, 91 (1982);
A. Aspect, J. Dalibard, and G. Roger,
Phys. Rev. Lett. \textbf{49}, 1804 (1982).



\bibitem{Weihs98}
G. Weihs \emph{et al.}, 
Phys. Rev. Lett. \textbf{81}, 5039 (1998).

\bibitem{Tittel98}
W. Tittel \emph{et al.}, 
Phys. Rev. A \textbf{57}, 3229 (1998).

\bibitem{Rowe01}
M.A. Rowe \emph{et al.}, 
Nature \textbf{409}, 791 (2001).




\bibitem{Pearle70}
Philip M. Pearle, Phys. Rev. D \textbf{2}, 1418 (1970).


\bibitem{Santos92}
E. Santos,  Phys. Rev. A \textbf{46}, 3646 (1992).





\bibitem{Polzik92}
E.S. Polzik, J. Carri, and H.J. Kimble,
Phys. Rev. Lett. \textbf{68}, 3020 (1992).



\bibitem{zhang03}
T.C. Zhang \emph{et al.}, 
Phys. Rev. A {\bf 67}, 033802 (2003).

\bibitem{Grosshans03}
F. Grosshans \emph{et al.}, 
Nature (London) \textbf{421}, 238 (2003).



\bibitem{Gilchrist98}
A. Gilchrist, P. Deuar, and M.D. Reid,
Phys. Rev. Lett. \textbf{80}, 3169 (1998).


\bibitem{Munro99}
W.J. Munro,  Phys. Rev. A \textbf{59}, 4197 (1999).

\bibitem{Wenger03}
J. Wenger \emph{et al.}, 
Phys. Rev. A \textbf{67}, 012105 (2003).


\bibitem{Ou92}
Z.Y. Ou \emph{et al.}, 
Phys. Rev. Lett. \textbf{68}, 3663 (1992).

\bibitem{Schori02}
C. Schori, J.L. S\o rensen, and E.S. Polzik,
Phys. Rev. A \textbf{66}, 033802 (2002).

\bibitem{Bowen04}
W.P. Bowen \emph{et al.}, 
Phys. Rev. A \textbf{69}, 012304 (2004).




\bibitem{Opatrny00}
T. Opatrn\'{y}, G. Kurizki, and D.-G. Welsch,
Phys. Rev. A \textbf{61}, 032302 (2000).

\bibitem{Cochrane02}
P.T. Cochrane, T.C. Ralph, and G.J. Milburn,
Phys. Rev. A \textbf{65}, 062306 (2002).

\bibitem{Olivares03}
S. Olivares, M.G.A. Paris, and R. Bonifacio,
Phys. Rev. A \textbf{67}, 032314 (2003).

\bibitem{WBG}
J. Wenger, R. Tualle-Brouri, and P. Grangier,
quant-ph/0402192; to appear in Phys. Rev. Lett. (2004).


\bibitem{CHSH} J.F. Clauser {\it et al.}, 
Phys. Rev. Lett. \textbf{23}, 880 (1969)

\bibitem{Bell} J.S. Bell, Speakable and Unspeakable in Quantum Mechanics
(Cambridge University Press, 
1988) p. 105.


\bibitem{Aspect76}
\noindent
A. Aspect, Phys. Rev. D {\bf 14}, 1944 (1976).


\bibitem{Sanchez04}
R. Garc\'{\i}a-Patr\'{o}n S\'{a}nchez, J. Fiur\'{a}\v{s}ek, and N.J.
Cerf, in preparation.


\bibitem{Eisert02}
J. Eisert and M.B. Plenio, Phys. Rev. Lett. \textbf{89}, 097901 (2002);
J. Fiur\'{a}\v{s}ek, Phys. Rev. A \textbf{66}, 012304 (2002).


\bibitem{Lam99}
P.K. Lam \emph{et al.}, 
J. Opt. B: Quantum Semiclass. Opt. \textbf{1}, 469 (1999).


\end{thebibliography}
\end{document}